# Non-resonant excitation of a two-level atom driven by a superoscillating field

D. G. Baranov,[1,2] A. P. Vinogradov,[1,2,3] and A. A. Lisyansky[4]

[1]*Moscow Institute of Physics and Technology, 9 Institutskiy per., Dolgoprudny 141700, Russia*
[2]*All-Russia Research Institute of Automatics, 22 Sushchevskaya, Moscow 127055, Russia*
[3]*Institute for Theoretical and Applied Electromagnetics, 13 Izhorskaya, Moscow 125412, Russia*
[4]*Department of Physics, Queens College of the City University of New York, Queens, NY 11367, USA*

We study counterintuitive dynamics of a two-level system (TLS) interacting with electric field superoscillating in time. We show that a TLS may be excited by an external light pulse whose spectral components are below the absorption line of the TLS. We attribute this unique dynamics to the Rabi oscillations of the TLS in a superoscillating driving field.

As was shown by Berry [1] in 1994, a band-limited function $f(t)$, i.e. the function whose Fourier transform satisfies $\hat{f}(\omega) = 0$ for all frequencies $|\omega| > \omega_{max}$, may oscillate with a frequency greater than $\omega_{max}$. Such functions were called superoscillatory (SO) functions. Since then, mathematical properties of SO functions have been studied in detail [2-4] and various approaches to their construction have been suggested [5,6]. Superoscillations have also been discussed in the context of quantum mechanical wavefunctions [7-9]. In particular, it was predicted [7] that a particle with a low momentum obtains a large momentum after passing through a slit if the superoscillating region of the wavefunction overlaps with the slit. SO field patterns were employed to construct field distributions with subwavelength resolution [10-15] without the use of near field. A microscope with the subwavelength resolution operating in the far field has also been demonstrated [16]. Some similarities between the phenomenon of superoscillations and speckle patterns can be established [17]. In all these studies, *spatial* oscillations are the object of interest. At the same time, the mathematical descriptions of SO functions are identical in both spatial and *time* domains, so that one may consider responses of different physical systems to superoscillating external perturbations. An example of such interactions, which we study here, is the dynamics of a TLS placed in an external SO electromagnetic field.

An atom in its ground state subjected to an oscillating electric field, it can be excited if the frequency of the oscillating field lies within the narrow absorption band of the atom $\omega_0 - \Gamma < \omega < \omega_0 + \Gamma$. Otherwise, the driving field interacts weakly with the atom unless the high intensity of the field causes multiphoton electronic transition. We address the intriguing question whether an atom can be excited by a superoscillating electromagnetic field, whose spectral components are outside of the absorption band of the atom.

In this Letter, we study the dynamics of a TLS, which models a real emitter, driven by a strong superoscillating electric field. We show that, in principle, by proper tailoring of driving field the TLS can be inversely populated by an external optical pulse which all spectral components of the electric field fall below the TLS absorption band. We show that this phenomenon cannot be associated with multiphoton absorption.

The simplest way to construct a SO function is to directly synthesize a quickly varying (in terms of maxima and minima repetition rate) function from low frequency components (see, e.g., Ref. [14]). The SO function is sought in the form of a superposition of $N$ harmonic oscillations:

$$s(t) = \text{Re} \sum_{n=1}^{N} c_n g_n(t), \quad g_n(t) = \exp(i\omega_n t) \tag{1}$$

where $|\omega_n| < \omega_{max}$. To obtain SO behavior for this function, we choose a sequence of times $t_n$, $n = 1...N$, to which certain values $s(t_n) = s_n$ are assigned. Then coefficients $c_n$ are obtained by solving the system of linear equations:

$$s(t_j) = \sum_{n=1}^{N} c_n \exp(i\omega_n t_j) = s_j \tag{2}$$

If the values $s_n$ are chosen such that they oscillate faster than the harmonic with the maximum frequency, $\omega_{max}$, then the function $s(t)$ will show SO behavior in the time interval $t_1 < t < t_N$.

In order to reveal the mechanism of the excitation process of an atom in the SO field, we study the dynamics of the system during a short pulse. To do this, we multiply the SO function by the exponential factor which results in shaping of the pulse:

$$f(t) = s(t) \exp\left(-\frac{(t-t_0)^2}{T^2}\right), \tag{3}$$

where $t_0$ is the center of the pulse and $T$ denotes its characteristic duration. Of course, considering a slowly-varying pulse instead of a harmonic oscillation inevitably causes broadening of the signal spectrum. As a consequence, a certain part of the $f(t)$ spectrum overlaps with the absorption band of the atom. However, as we show below, a non-resonant excitation of the atom cannot be attributed to this broadening, provided that the pulse duration, $T$, is large enough.

The object of our interest is a quantum emitter driven by an external oscillating electric field. We model the atom as a TLS with ground and excited states denoted by $|g\rangle$ and $|e\rangle$, respectively. The dynamics of a driven TLS is governed by the time-dependent Schrödinger equation with the usual Hamiltonian

$$\hat{H}_0 = \hbar\omega_0 \hat{\sigma}^\dagger \hat{\sigma} - \Omega f(t)(\hat{\sigma}^\dagger + \hat{\sigma}), \qquad (4)$$

where $\omega_0$ is the TLS transition frequency, $\hat{\sigma} = |g\rangle\langle e|$ and $\hat{\sigma}^\dagger = |e\rangle\langle g|$, $\Omega = \mu E$ is the Rabi constant describing coupling of the atom and the electric field, and $f(t)$ is the real-valued electric field shape function. Below we measure all frequencies in the units of $\omega_0$ and all times in the units of the inverse transition frequency $\omega_0^{-1}$. To take spontaneous decay into account, and dephasing of the atom, we use the density matrix operator $\hat{\rho}$. In the framework of the density matrix operator formalism, the equations governing the system evolution are [18]:

$$\begin{aligned}\dot{\rho}_1 &= \rho_2 - \rho_1/T_2, \\ \dot{\rho}_2 &= -\rho_1 - \rho_2/T_2 + 2\Omega f(t)\rho_3, \\ \dot{\rho}_3 &= -2\Omega f(t)\rho_2 - (\rho_3 - \rho_{30})/T_1,\end{aligned} \qquad (5)$$

In Eqs. (5), terms $\rho_{1,2,3}$ are expressed through the elements of the density matrix as $\rho_1 = \rho_{12} + \rho_{21}$, $\rho_2 = i(\rho_{12} - \rho_{21})$ and $\rho_3 = \rho_{22} - \rho_{11}$. In order to take into account the effects of dissipation, the phenomenological diagonal terms corresponding to spontaneous decay and dephasing processes with characteristic times $T_1$ and $T_2$, respectively, are included into Eqs. (5). The term $\rho_{30}$ represents the population inversion of the TLS at rest and it can be assumed that $\rho_{30} = -1$ at room temperature, i.e., the atom decays to its ground state when not acted upon by an electric field.

To illustrate the dynamics of an atom in the SO field let us choose certain values for decay times and Rabi constant and numerically solve the dynamical Eqs. (5). Let the dimensionless decay and dephasing times be equal to $T_1 = 500$ and $T_2 = 300$, respectively, and the Rabi constant to $\Omega = 0.01$.

For components of the band-limited SO function $s(t)$ we choose a basis of five harmonic oscillations $g_n(t) = e^{i\omega_n t}$ with frequencies

$$\omega_n = 0.18n, \; n = 1,...,5. \qquad (6)$$

All harmonics from this set satisfy the condition $|\omega_n| < \omega_0$. With this basis of harmonic oscillations, we intend to obtain a SO function that oscillates locally during a certain time interval with the frequency of the atomic transition $\omega_0$, which is *larger than the maximum frequency presented in the spectrum*. To do this, we choose five points $t_n = \pi n / \omega_0$, $n = 0, ..., 4$, to which the following values are assigned: $s_1 = s_3 = s_5 = -1$, $s_2 = s_4 = 1$ [see the pattern depicted in Fig. 1(a)]. Solving the linear system of Eqs. (2), we obtain the desired SO function $s(t)$ which is plotted in Fig. 1(b). For the given pattern $s_n$, one obtains complex amplitudes for the harmonics:

$$c_1 \approx -0.156 + 0.331i, \quad c_2 \approx -0.862 - 1.042i,$$
$$c_3 \approx 2.341 - 0.601i, \quad c_4 \approx -0.502 + 2.634i, \quad (7)$$
$$c_5 \approx -1.820 - 1.322i.$$

Fig. 1(b) shows the SO function $s(t)$ and the harmonic having the highest frequency from the basis (6). One can see from this figure that $s(t)$ oscillates within the interval $0 < t < 4\pi$ with local frequency $\omega_{loc} = 1$.

Now, we investigate the dynamics of the atom in the external pulse of the form $f(t) = As(t)\exp\left[-(t-t_0)^2/T^2\right]$ with the pulse duration $T = 100$ and position of the pulse center $t_0 = 200$. Here $A$ is a real-valued constant amplitude whose value is chosen in such a way that strong coupling between atom and electric field is attained. The pulse shape and its spectral density $\hat{f}(v) = \left|\int f(t)e^{ivt}dt\right|$ are plotted in Fig. 1 (c) and (d). Five peaks in Fig. 1(d) represent five harmonic components of the SO function $s(t)$ which are broadened due to the exponential factor $\exp\left[-(t-t_0)^2/T^2\right]$.

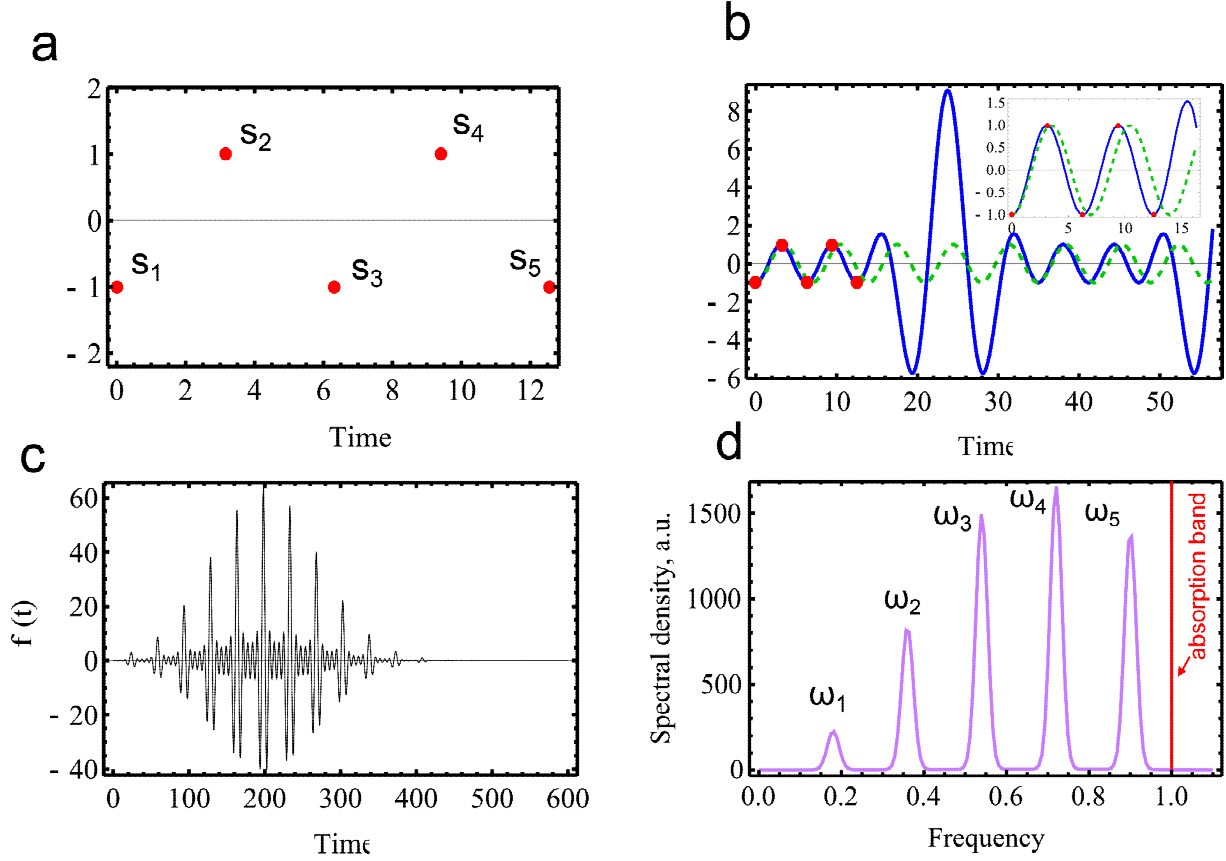

FIG. 1. (a) Predefined pattern of five points $\{t_n, s_n\}$ responsible for SO behavior of the function $s(t)$. (b) Solid line: the SO function $s(t)$ obtained by solving Eqs. (2) for the pattern depicted in panel (a). Dashed line: the fastest harmonic component of $s(t)$ having frequency $\omega_5 = 0.9$. The inset shows the time interval at which SO behavior is clearly visible. (c) The pulse shape function $f(t)$ with the duration $T = 100$ and the amplitude $A = 7$. (d) Normalized spectral density of pulse depicted in panel (c). Five peaks correspond to five spectral components of the initial SO function $s(t)$. Thin red line shows the absorption band of the atom defined as $\omega_0 - \Gamma < \omega < \omega_0 + \Gamma$, with $\omega_0 = 1$ and $\Gamma = 1/500$. Peaks of electric field pulse do not overlap with the absorption band.

We first present the dynamics of the atom driven by the field $f(t)$. The time dependence of the atom-field coupling $\Omega f(t)$ and the population inversion $\rho_3$ are shown in Fig. 2 for three different values of the pulse amplitude $A = 1$, $A = 4$ and $A = 7$ (we chose these values of external pulse amplitudes as they clearly illustrate the dynamics of the TLS).

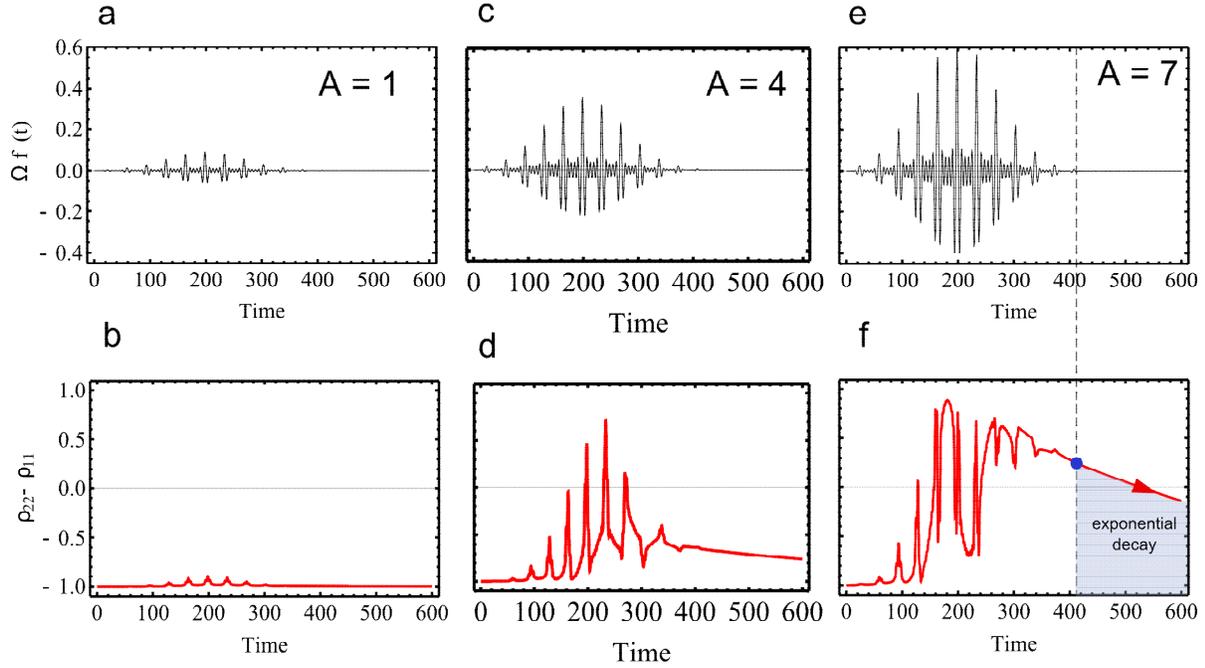

FIG. 2. (a, c, e) Time-dependent Rabi frequency $\Omega f(t)$ for a superoscillating pulse of three different amplitudes $A = 1, 4,$ and $7$. (b, d, f) Population inversion of the two-level atom driven by the field $f(t)$. For $A = 7$, the inverted population $\rho_{22} - \rho_{11} > 0$ is attained at the end of the superoscillating pulse.

Figure 2 represents the result central to our paper. When the atom-field coupling $\Omega f$ is small (left panels, $A = 1$), the atom almost does not interact with the external field [Fig. 2(b)]. At the middle panels, which correspond to the amplitude $A = 4$, one can observe population oscillations. Within the duration of the pulse, the population inversion may become positive, as indicated in Fig. 2(d). However, when the pulse action is finished, the population inversion returns nearly to its initial value $\rho_{30} = -1$, despite the superoscillating behavior of the pulse.

The right panel shows the most interesting result. In fact, it shows that a two-level atom can be excited by a non-resonant external electromagnetic field. At the end of the pulse, the atom is excited with the inverted population $\rho_{22} - \rho_{11} > 0$. When the pulse action is finished ($t > 400$), the atom experiences the usual exponential decay to its ground state. We claim that this unique non-resonant process of excitation is the direct manifestation of superoscilaltions of the external field.

Let us now illustrate the significance of superoscillations for observation of non-resonant excitation. Consider evolution of the atom population inversion in the electric field pulse $f(t) = A\cos(\omega_5 t)\exp\left[-(t-t_0)^2/T^2\right]$ formed by a single harmonic having the highest frequency $\omega_5$ from the set (6). The amplitude $A = 12$ of such a pulse is chosen in such a way that the characteristic Rabi frequency $\Omega f$ during the pulse is of the same order as for the SO pulse for which the inverted population is attainable. Results presented in Fig. 3 clearly indicate that use of a single non-resonant harmonic does not allow one to achieve a positive population inversion. At the end of the pulse (time position $t = 400$), the value of the population inversion is $\rho_3 \approx -0.7$. For even stronger pulses, the resulting atom population inversion is of the same order. No positive population inversion was observed in numerical simulation with a pulse of frequency $\omega_5$.

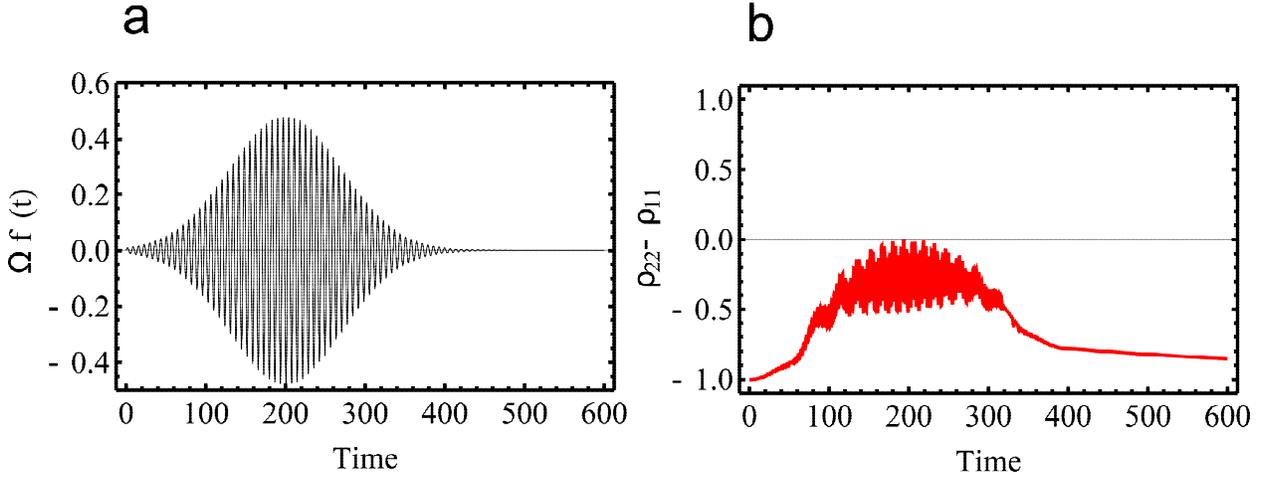

FIG. 3. Illustration of non-resonant light-atom interaction without superoscillations. (a) The Rabi frequency $\Omega f$ of a pulse formed by a single harmonic with frequency $\omega_5$. Characteristic value of the Rabi frequency is comparable to that observed in Fig. 1(e). (b) Corresponding dynamics of the atom population inversion $\rho_3 = \rho_{22} - \rho_{11}$.

The observed behavior of a two-level atom in a superoscillating pulse cannot be attributed to spectrum broadening. To justify this, we consider evolution of an atom in the pulse of the same duration $T = 100$ formed by the resonant harmonic of frequency $\omega_0$:

$$f_{res}(t) = A_{res}\,\mathrm{Re}\exp(i\omega_0 t)\exp\left(-\frac{(t-t_0)^2}{T^2}\right). \tag{8}$$

We set the value of amplitude $A_{res} = 0.02$. The spectral densities of such a resonant pulse and the SO pulse for which we observe inverted population, are shown in Fig. 4(a). Spectral densities of these two pulses are of the same order near the absorption region of the atom. In Fig. 3(b), we plot the corresponding dynamics of the atom population inversion coupled to the pulse of electric field (8). The resulting population inversion during the pulse is negligible in comparison with the values shown in Figs. 2 for the case of the SO pulse. This dependence clearly indicates that interaction of the atomic transition with the resonant part of the spectrum due to broadening is completely insufficient for achieving pronounced population inversion. Therefore, the observed dynamics of the TLS can only be associated with the superoscillating behavior of electric field.

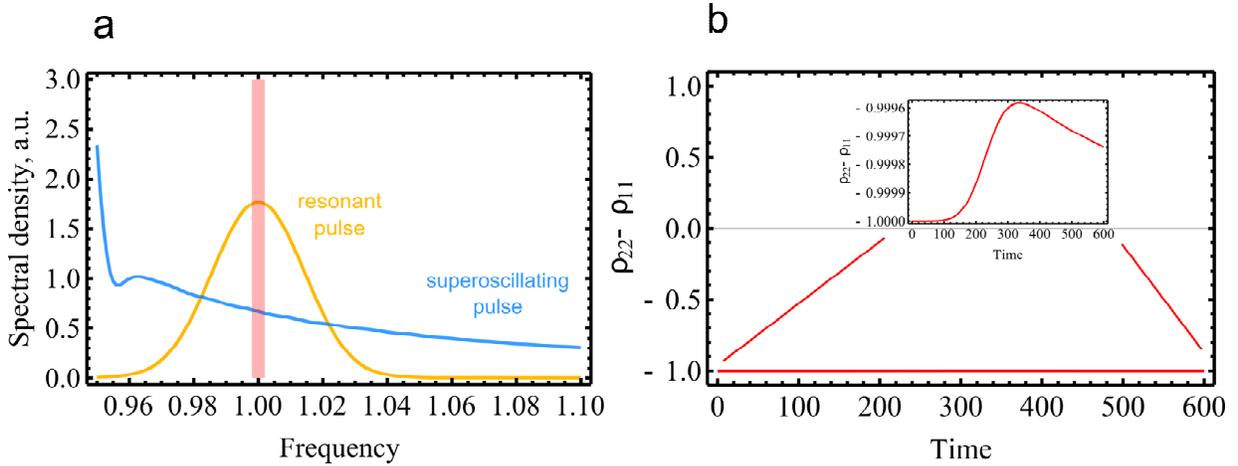

FIG. 4. Dynamics of a two-level atom in the resonant electric field pulse. (a) Spectral densities of the SO pulse and the resonant pulse. The red shaded region indicates the absorption band of the atom. (b) Corresponding dynamics of the two-level atom interacting with the resonant pulse $f_{res}(t)$.

It is worth emphasizing that, there are some fundamental limitations on superoscillations, which could make their use problematic in practice [3,19]. The main obstacle in the construction of a SO function is that the $L_2$ – norm of the resulting function increases polynomially with the desired frequency of superoscillations and exponentially with the number of oscillations. As a result, the amplitude of high frequency oscillations becomes small in comparison with that of the low-frequency region. In other words, the energy of a SO signal is concentrated in the lower frequency region.

It may seem that a high intensity field may cause the multiphoton excitation of the TLS. We show that this is not so. The phenomenon demonstrated here cannot be explained by a multiphoton absorption. Frequencies $\omega_n$ making up the basis of harmonic oscillations $g_n(t)$ in Eq. (1) are arbitrary. The only condition which the basis set $g_n(t)$ *must* satisfy is the solvability of Eqs. (2). Thus, we can choose such a set of frequencies that for any pair of frequencies $\omega_k$ and $\omega_l$ the condition of the two-photon absorption $\omega_k + \omega_l = \omega_0$ *does not* hold. This is the case for the particular set of frequencies (6) that we use in present calculations.

The TLS studied here is a great simplification of the energy structure of a real atom. Atomic structures may contain transitions whose frequencies are close to components $\omega_n$ of the SO electric field. Therefore, those transitions will strongly interact and absorb lower-frequency photons, which play crucial role in superoscillating behavior of the electric field. Nevertheless, as noted in the previous paragraph, the basis frequencies are to some extent arbitrary. Thus, it allows one to tailor the set of frequencies $\omega_n$ in such a way that *none of these frequencies* is attributed to absorption bands of a certain energy structure of an atom.

In conclusion, we have investigated the unique temporal dynamics of a two-level atom driven by an external superoscillating field. It is possible to construct a superoscillating pulse of electromagnetic field, using only non-resonant harmonics, which can excite the atom. By tailoring the shape and amplitude of the pulse we achieve population inversion of the atom at the end of the superoscillating pulse. The observed dynamics is associated with the superoscillations phenomenon and cannot be attributed to multiphoton processes or spectrum broadening. This phenomenon has clear application to exciting atomic transitions by a laser pulse with a spectrum falling below the atomic resonances.

The work was supported by RFBR grants No 13-02-00407 and 13-07-92660 and by the NSF under Grant No. DMR-1312707